# Convolutional LSTM Surrogate for Mesoscale Hydrocode Simulations of Granular Wave Propagation

Kathleen Winona Vian Martinus[1], Sushan Nakarmi[1],

Dawa Seo[2], Nitin Pandurang Daphalapurkar[1]*

[1] Theoretical Division

Los Alamos National Laboratory

P.O. Box 1663, Los Alamos, NM 87545

[2] Civil and Environmental Engineering

Wayne State University, 5050 Anthony Wayne Dr, Detroit, MI 48202

**Abstract**

Granular materials subjected to impact loading exhibit highly heterogeneous spatiotemporal dynamics governed by wave propagation, pore collapse, and grain-scale rearrangements. Mesoscale hydrocodes resolve these processes but are computationally expensive, limiting their use in parametric studies and uncertainty quantification. In this work, we develop a convolutional Long Short-Term Memory (ConvLSTM) neural network as a spatiotemporal surrogate for mesoscale simulations of weak shock propagation in granular media. Using two-dimensional hydrocode simulations as training data, we first consider a simplified "billiard break" problem in which a cue ball impacts a cluster of nine circular balls, all deformable. Sequences of pressure-field images

Corresponding author: nitin@lanl.gov

serve as input–output pairs for a sequence-to-sequence ConvLSTM, which is trained to predict future frames from a short history. We compare several architectures and show that a relatively compact encoder–decoder ConvLSTM accurately reproduces the propagation of the pressure wave and the resulting particle motion for an unseen combination of cue-ball position and impact velocity. As a proof-of-concept extension, we apply the same ConvLSTM framework to previously published mesoscale simulations of weak shock compaction in a granular ensemble. When evaluated at piston impact speeds that were completely withheld from training, the surrogate captures the position and shape of the compaction front and its dependence on impact speed, while smoothing fine pore-scale details in the highly compacted region as expected. These results demonstrate that ConvLSTM models can serve as satisfactory surrogates for spatiotemporal mesoscale simulations of granular wave propagation, enabling accelerated exploration of parameter space and laying the groundwork for physics-informed, mesoscale simulations of granular materials under shock loading.

# 1 Introduction

Granular materials under dynamic loading arise in a broad spectrum of applications, including powder metallurgy, mining, deformation of energetic materials, and crater formation [1, 2]. Their response is governed by complex interactions between grains, pores, and propagating stress waves, leading to heterogeneous fields of pressure, strain, and velocity at the mesoscale. Resolving these mechanisms is essential for reliable continuum-scale models, but direct experimentation often lacks the spatial and temporal resolution needed to fully characterize statistically representative pore-scale dynamics [3].

Mesoscale numerical simulations, in which individual grains are modeled as deformable solids with explicit pore structures, have emerged as a powerful tool to bridge the gap between grain-scale physics and bulk response [1, 4]. Recent work using Los Alamos National Laboratory's (LANL) Free Lagrange (FLAG) hydrocode has demonstrated the importance of pore-size distribution (PSD), particle shape, and initial contact networks for weak shock com-paction in granular salt [5, 2]. These studies show that shock compaction velocity, wavefront thickness, and internal stress substructure are highly sensitive to particle-scale features and that mesoscale modeling provides pore-scale insights inaccessible to traditional continuum descriptions.

However, high-fidelity mesoscale simulations are computationally expensive. Each simulation required resolving on the order of ten thousand grains in 2D and fine temporal resolution over many time steps. Consequently, routine tasks such as parametric sweeps, design optimization, and uncertainty quantification become prohibitive when relying exclusively on direct mesoscale computations. This motivates the search for *surrogate models* that approximate the behavior of the mesoscale solver but are sufficiently fast to evaluate repeatedly.

Machine learning (ML) offers a natural framework for such surrogates. Recent reviews have documented the rapid growth of ML methods in granular and geomechanical modeling [6, 7]. Applications include neural constitutive models, encoder–decoder architectures, graph neural networks (GNNs), neural operator methods, and uncertainty quantification. In granular mechanics, GNN-based surrogate models have been shown to accelerate discrete element method (DEM) simulations of granular flows by orders of magnitude while retaining accuracy [8]. Deep active learning and transfer learning strategies have been used to build data-driven constitutive models from DEM simulations of granular soils [9, 10, 11]. Recurrent architectures such as Long Short-Term Memory (LSTM) and gated recurrent unit (GRU) networks have been applied to learn stress–strain histories and path-dependent responses in granular materials [12, 13, 14].

Despite this progress, most existing ML models for granular materials operate on low-dimensional inputs and outputs, such as scalar stress and strain histories or averaged quantities. Full-field spatiotemporal modeling of pressure, velocity, and grain configuration remains largely unexplored. In particular, there is limited work on learning the co-evolution of spatially heterogeneous pressure fields and grain-scale kinematics directly from mesoscale continuum simulations.

In this paper, we develop a spatiotemporal surrogate model for mesoscale granular wave propagation based on a convolutional Long Short-Term Memory (ConvLSTM) neural network [15]. ConvLSTM extends the classical LSTM by replacing dense matrix multiplications with convolutions, enabling it to capture spatial correlations in image sequences while maintaining temporal memory. We train the ConvLSTM on sequences of images rendered from FLAG hydrocode simulations, where color encodes the pressure field and grain positions in a two-dimensional domain.

We first focus on a simplified "billiard break" problem in which a cue ball impacts a cluster of nine circular balls. This system provides a controlled setting for developing and testing the methodology. We vary the cue-ball position and velocity to generate a family of mesoscale trajectories and use them to train and test the ConvLSTM. We then apply the same architecture to mesoscale simulations of weak shock compaction in a granular ensemble of circular grains from Seo et al. [5], using piston impact speed as a control parameter. The compaction problem is treated as a proof-of-concept extension to assess whether the ConvLSTM trained on mesoscale data can interpolate across impact speeds and reproduce the large-scale features of the compaction front.

The remainder of the paper is organized as follows. Section 2 summarizes related work on ML for granular materials and LSTM-based approaches for path-dependent behavior. Section 3 describes the FLAG hydrocode simulations, the image-based data representation, and the ConvLSTM architecture and training procedure. Section 4 presents results for the billiard problem and the granular compaction extension. Section 5 discusses implications, limitations, and opportunities for future work. Conclusions are given in Section 6.

# 2 Background

Recent advances in machine learning have enabled novel modeling approaches for granular materials across scales. A comprehensive review by Wang et al. [6] and the references therein categorizes ML applications at both the grain scale and the bulk continuum scale, including neural constitutive modeling, encoder–decoder networks, graph learning, neural operator methods, and uncertainty quantification. This work identifies growing interest in integrating physical priors and domain knowledge into data-driven methods.

Fransen et al. [7] discuss key challenges for applying scientific machine learning to granular simulations, particularly at the continuum scale. Their position paper highlights the importance of sequence models, graph-based representations, and probabilistic surrogates for capturing the complex, multiscale nature of granular mechanics.

Choi and Kumar [8] demonstrate the potential of graph neural networks (GNNs) to serve as surrogate models for granular flows such as column collapses. Their graph neural simulator (GNS) achieves orders-of-magnitude speedup compared to traditional simulation while maintaining high predictive accuracy. Qu et al. [9] proposed a deep active learning framework for modeling granular material response using Discrete Element Method (DEM) data. In related work, Qu et al. [10] apply knowledge transfer techniques for multiscale modeling, enabling efficient training from synthetic datasets and improving generalization across loading regimes.

Several publications have explored the use of recurrent neural networks (RNNs) and, in particular, Long Short-Term Memory (LSTM) networks for modeling the path-dependent mechanical response of granular materials. A study on deep transfer learning-aided constitutive modeling [11] trained LSTM models on DEM simulations, capturing complex stress-strain relationships under cyclic and monotonic loads. Zhang et al. [12] demonstrated that both LSTM and GRU networks perform effectively in learning stress–strain behavior from DEM-based triaxial tests. To address history dependence, Ma et al. [13] introduced a modified LSTM architecture in which the initial hidden state encodes the granular material's physical state (e.g., density, stress), enabling accurate predictions under non-monotonic loading paths. Additional work on LSTM-based surrogates has focused on modeling granular flow fields and their evolution in time [14], showing promising results in replacing time-expensive DEM simulations. Ulloa et al. [16] extend model-free data-driven mechanics to micromorphic continua, enabling simulations of strain-localizing/softening materials that capture the intrinsic material length scale and associated instabilities without prespecifying a failure mode.

Mesoscale simulations for shock compaction of granular materials have advanced significantly. Vogler et al. [1] and others used mesoscale finite element simulations to explore differences between two- and three-dimensional (2D and 3D) formulations and the effects of contact laws on dynamic compaction. More recently, Seo and co-workers conducted systematic mesoscale studies of weak shock compaction in granular salt (square grain morphology), focusing on pore-size distribution (PSD) effects [5] and related configurations [2]. These works showed that the shock compaction velocity and front structure depend strongly on PSD, particle shape, and initial contact configurations, and introduced statistical metrics such as coordination number, mean pore diameter, and bimodality coefficient to characterize PSD.

Despite the breadth of these contributions, most prior LSTM/GRU models for granular materials have commonly focused on low-dimensional inputs. The co-evolution of field-level pressure and particle-level positions is a unique, multiphysics ML target that is not handled in existing literature. Likewise, ML modeling of granular materials via continuum solvers, especially with elastodynamic fidelity (e.g., accounting for shocks, reflections, and wave interference), remains largely unexplored. Spatially aware LSTM architectures such as ConvLSTM have not been widely applied to granular material modeling, particularly for continuum-level fields derived from elastodynamic hydrocodes. This work bridges that gap by combining high-fidelity mesoscale simulations with ConvLSTM-based surrogates to learn the spatiotemporal evolution of pressure fields and grain configurations.

# 3 Methods

## 3.1 Mesoscale Modeling using FLAG hydrocode

Our prior work developed a mesoscale computational model to simulate the shock compaction test with square-shaped grains using the FLAG [17, 18] hydrodynamic code. The Lagrangian method was preferred because interface tracking is natural to this method, and physically based contact models can be employed. Since impact velocities were in a weak shock regime, we did not anticipate any issues related to mesh distortion. In this work, the simulations employed circular grains within a two-dimensional simulation setting while retaining the mechanical properties of salt grains. Frictional forces, both the friction between grains and the friction between the grain and tube walls, were not accounted for in the model.

Discretization involved creating a mesh of a four-node element (aka a zone) within a material domain occupied by a particle. FLAG then solves for field variables and thermodynamic quantities using material constitutive and equation of state continuum models accounting for normal contact between the deformable balls. Mesh convergence studies were performed to select a zone size that resolves stress gradients associated with small-scale features while balancing computational cost [2]. However, using more zones for higher mesh resolution can be computationally expensive. Therefore, it was essential to determine a reasonable mesh resolution that balances the numerical solution's accuracy and efficiency.

In this study, we used an elastic-plastic response with linear hardening, along with the equation of state referencing the SESAME (Salt: 7282, Sapphire: 7411, Brass: 4100) ([19]). Each simulation was performed using high-performance computing at LANL, utilizing 144 processors across four nodes spanning 12 hours. The mesoscale procedure was previously validated against the measured $U_s$-$U_p$ response for granular salt, as reported in Seo et al. (2024)[2]. We applied the same testing setup (see Figure 14 in Ref. [2] ) and material properties in this work as in Seo et al. (2024)[2] (Table 1).

| Materials | Salt[20] | Brass[21, 22] | Sapphire[23, 24] |
|---|---|---|---|
| Density (g/cm$^3$) | single grain: 2.16;sample: 1.41 | 8.47 | 3.97 |
| Shear modulus (GPa) | 10.48 | 38.7 | 175 |
| Hardening modulus (GPa) | 2.34 | 20 | 20 |
| Yield strength (MPa) | 11.8 | 337 | 380 |
| SESAME table | 7282 | 4100 | 7411 |

Table 1: Material Properties for mesoscale simulations

## 3.2 Mesoscale hydrocode simulations

### 3.2.1 Billiard impact problem

We first consider a simplified two-dimensional "billiard break" configuration consisting of nine circular balls in contact, impacted by a single "cue ball" at varying positions and velocities. All simulations were performed with the FLAG hydrocode, using the same material model, equation of state, and boundary conditions as in Seo et al. [5]. The billiard problem is treated as a mesoscale elastodynamic system in which the impacting cue ball generates a transient stress wave traversing through individual grains, that then propagates through the particle assembly and induces particle motion.

In total, twelve distinct FLAG simulations of the 10-ball billiard configuration were carried out. Each simulation differed in the *initial position* and *velocity vector* of the cue ball. The cue ball was initialized at three locations along the lower boundary of the domain,

$$(x_0, y_0) \in \{(0.0, -4.0),\ (1.0, -4.0),\ (-1.0, -4.0)\} \text{ cm}, \tag{1}$$

and launched with one of four velocity vectors,

$$(v_x, v_y) \in \{(0.003, 0.02),\ (0.003, 0.03),\ (0.002, 0.02),\ (0.002, 0.03)\} \text{ cm}/\mu\text{s}. \tag{2}$$

Changing the horizontal and vertical components of the velocity simultaneously varied both the impact speed and the incident angle of the cue ball, as well as the overall momentum input into the granular assembly.

For the billiard case, the FLAG time integration used a maximum time step of $\mathrm{d}t_{\max} = 0.025\ \mu$s and a final time $t_{\text{stop}} = 2000\ \mu$s. Data including pressure and particle positions were written every 70 time steps and post-processed into two-dimensional color images. This output resulted in 1,152 images per simulation and 13,824 total images between the twelve simulations.

Of the twelve simulations, ten were used for training the ConvLSTM model and two were reserved exclusively for testing. The test simulations correspond to

- Test set 1: $(v_x, v_y) = (0.002, 0.03)$ cm/$\mu$s, cue-ball position $(1.0, -4.0)$ cm,
- Test set 2: $(v_x, v_y) = (0.002, 0.03)$ cm/$\mu$s, cue-ball position $(-1.0, -4.0)$ cm.

The individual velocities and positions appearing in these test cases are present in the training set, but the specific *combinations* of position and velocity of the cue ball used in the test runs are not. This design probes the model's ability to interpolate between previously seen initial conditions without requiring extrapolation to completely unseen parameter ranges. A validation set was constructed by randomly withholding 10% of the training sequences (at the sequence level), using a validation fraction of 0.1 in the data loader.

### 3.2.2 Granular compaction problem

As a proof-of-concept extension beyond the simplified billiard system, we also applied the ConvLSTM framework to weak shock compaction of a dense granular medium. These simulations follow the mesoscale setup of Seo et al. [5], in which circular grains represent a granular solid impacted by a rigid piston. The material model (SESAME equation of state for NaCl and elastic-plastic constitutive response with linear hardening), boundary conditions (rigid piston, lateral confinement, and particle–wall interactions), and overall numerical settings are identical to those used in that prior work.

In the full granular compaction simulations, the total number of finite volume zones (equivalent to finite elements) in the domain is on the order of several thousand. For training the ConvLSTM surrogate, we extracted a focused sub-window of the domain containing several hundred grains (fewer than one thousand) near the compaction front. This cropped region captures the evolving shock front, particle rearrangements, and porosity evolution while keeping the image resolution compatible with the ConvLSTM input size.

We ran ten granular compaction simulations with eight used for training the ConvLSTM model and two reserved exclusively for testing. A validation set was constructed by randomly withholding 10% of the training sequences at the sequence level.

The only parameter varied across simulations was the piston impact speed, which was sampled between 50 m/s and 400 m/s. All simulations with piston speeds in this range, excluding 100 m/s and 300 m/s, were used for training the ConvLSTM. The simulations at 100 m/s and 300 m/s were completely withheld from training and served as independent test cases to assess the surrogate's ability to interpolate across impact speeds in a more complex granular setting. We use the following impact velocities in our simulations:

$$v \in \{50,\ 100,\ 150,\ 200,\ 250,\ 275,\ 300,\ 350,\ 375,\ 400\} \times 10^{-4}\ \text{cm}/\mu\text{s}. \tag{3}$$

FLAG simulations were advanced with a maximum time step of $dt_{max} = 0.001\ \mu s$ and terminated at a final time of 15 $\mu$s. Output was written at fixed physical-time intervals of 0.01 $\mu$s and post-processed into two-dimensional color images. This time-based output produced in 1,501 images per simulation, for a total of 15,010 images across the ten granular compaction simulations.

## 3.3 Image representation and sequence construction

The surrogate is trained to learn the rendered field rather than raw pressure values. For both the billiard and granular compaction problems, FLAG field outputs were converted into two-dimensional color images. Each saved frame was written to disk as a 128 × 128 RGB image, where color encodes both the pressure field and the configuration of particles. The exact mapping of RGB values, encoding both pressure and geometric information, is determined by the visualization pipeline implemented in the open-source ParaView software package.

Images were loaded using TensorFlow's utility, which returns tensors of shape $(128, 128, 3)$ with integer pixel intensities in the range $[0, 255]$. For each image, we convert the RGB values to floating point and normalize each channel by dividing by 255, resulting in pixel values scaled to the range $[0, 1]$. Because the particle shapes, positions, and pressure magnitudes are all encoded in the RGB color values, no additional binary masks or explicit particle labels are required; the spatial structure of the granular assembly is implicitly represented in the color image.

We adopt a sequence–based, sliding-window formulation that enables the ConvLSTM model to learn temporal dependencies across image frames. We define overlapping input sequences of ten consecutive frames,

$$\mathbf{x}_t = [f_t, f_{t+1}, \ldots, f_{t+9}], \tag{4}$$

which are extracted from each simulation and mapped to corresponding target sequences

$$\mathbf{y}_t = [f_{t+1}, f_{t+2}, \ldots, f_{t+10}], \tag{5}$$

where $f_t$ denotes the image at frame index $t$. During training, these sequences are incrementally shifted forward in time, allowing temporal overlap between samples and increasing the effective size of the training set. The model is optimized to predict multiple future frames from each input sequence, with the loss evaluated only on the terminal prediction. This promotes stable learning of multi-step temporal dynamics and avoids the instability associated with fully autoregressive rollouts.

During inference, the trained model is initialized using only the first ten frames and applied iteratively in a sliding-window manner. Each prediction step produces a 10-frame sequence of future timesteps, and only the terminal frame from that sequence is retained and used to construct the next input window. As a result, feedback occurs at the level of multi-step evolution rather than through direct one-step recursion, which reduces error accumulation and promotes more stable long-horizon predictions.

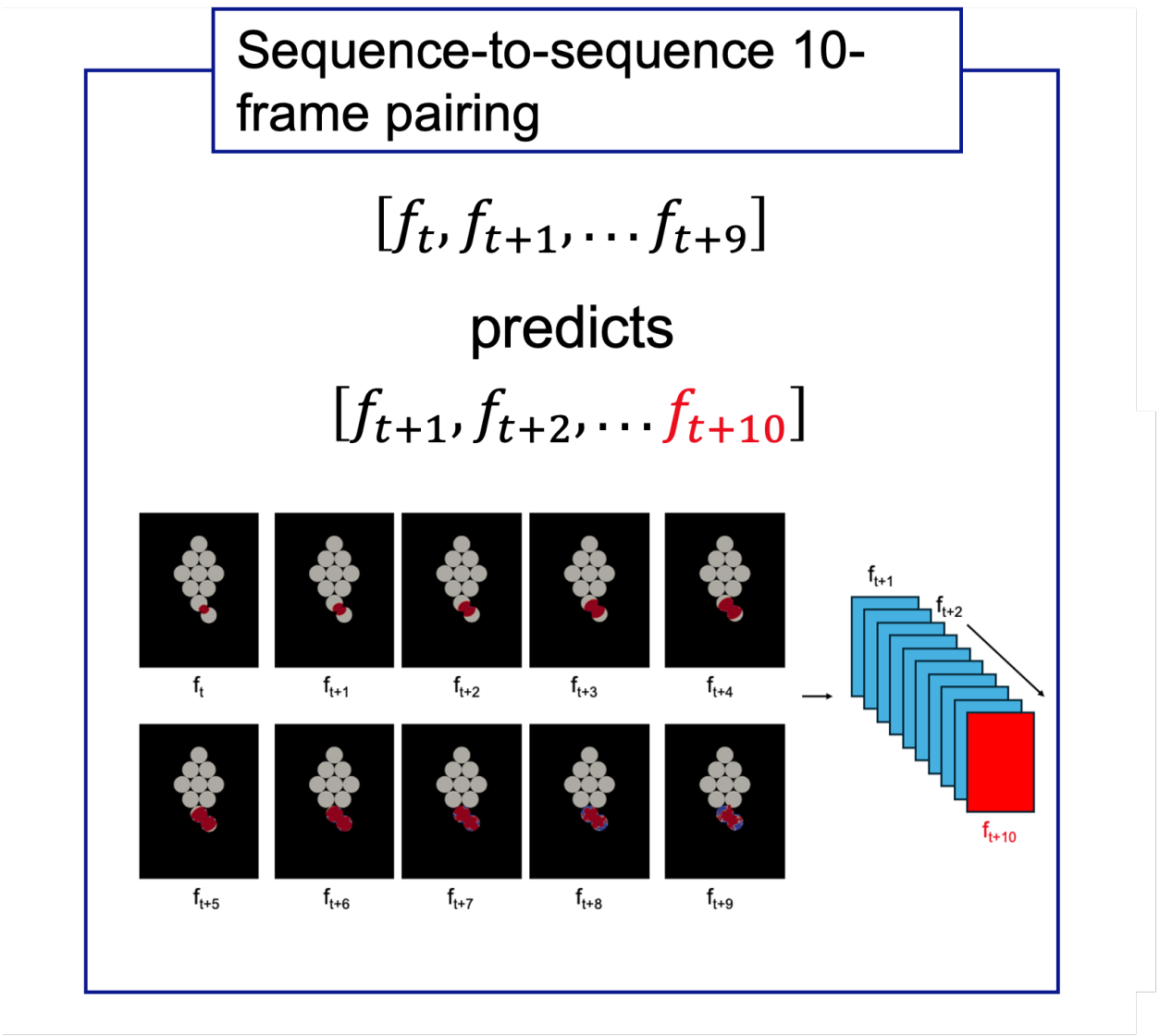

Figure 1: Sliding window approach during training

We note that while neural-operator methods are a powerful alternative for learning parametric solution maps of PDEs, they are most naturally applied to problems with clearly defined field variables on consistent discretizations and with ample coverage of the parameter space. In contrast, our training data consist of rendered RGB frames that jointly encode pressure and evolving particle geometry (interfaces, contacts, and topology changes) in a strongly transient, weak-shock regime. Given the limited number of expensive mesoscale trajectories available for training and the presence of sharp moving wavefronts and contact-driven discontinuities, we adopt a ConvLSTM architecture as a pragmatic and robust image-sequence surrogate. ConvLSTM directly models spatiotemporal correlations in the rendered fields without requiring explicit interface tracking or operator inputs, and provides stable multi-step predictions under our sliding-window training strategy. Neural operators remain a promising direction for future work, particularly as larger parametric datasets and explicit field representations become available.

## 3.4 Convolutional LSTM architecture

We adopt an encoder–decoder ConvLSTM architecture that captures both local spatial structure and temporal evolution in the image sequences. ConvLSTM replaces the fully connected operations in a standard LSTM with convolutions, enabling the network to maintain spatially structured hidden states [15]. In each ConvLSTM layer, the input-to-state and state-to-state transitions are implemented as convolutions, and the gating mechanisms (input, forget, and output gates) operate on feature maps rather than vectors.

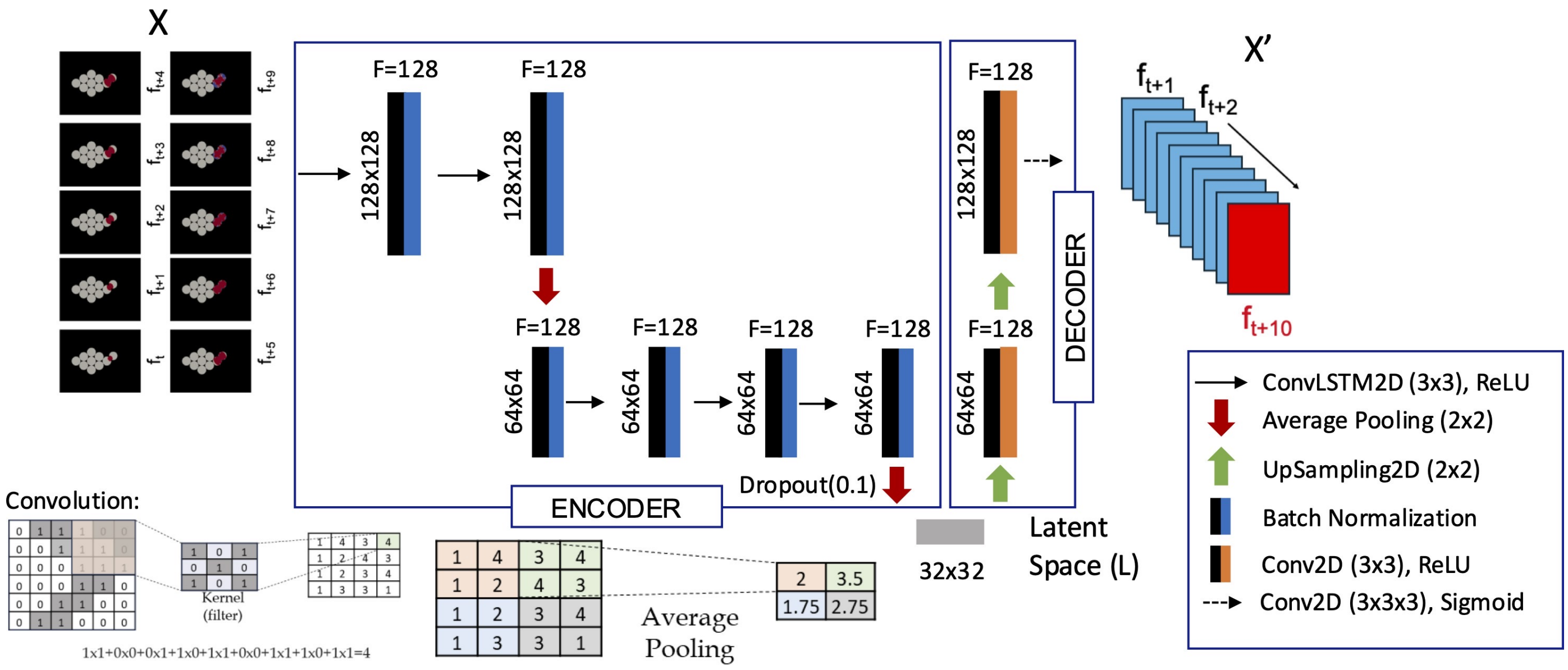

Figure 2: Schematic of the encoder–decoder ConvLSTM architecture used in this work, showing ConvLSTM layers, pooling/upsampling operations, and the sequence-to-sequence mapping from input frames to predicted frames.

The encoder consists of a stack of ConvLSTM layers with increasing receptive field, interleaved with spatial downsampling. In the configuration used here, six ConvLSTM layers are applied in sequence, each with 128 filters and 3 × 3 kernels and rectified linear unit (ReLU) activations. After the second and fourth ConvLSTM layers, we apply average pooling to reduce the spatial resolution, yielding a latent representation on a 32 × 32 grid. This gradual reduction in spatial resolution allows the network to aggregate information over larger spatial neighborhoods while controlling the number of parameters.

The decoder mirrors the encoder structure and reconstructs the full-resolution image sequence from the latent representation. It consists of two ConvLSTM layers with 128 filters and ReLU activations, followed by upsampling operations that restore the spatial resolution to 128 × 128, matching the input size. A final ConvLSTM layer with a sigmoid activation produces the output sequence, with three channels corresponding to the RGB intensities at each pixel.

We explored several architectural variants, including different numbers of layers, kernel sizes (3 × 3 vs. 5 × 5), numbers of filters per layer, and pooling strategies Table 2. These variants were compared using mean squared error (MSE) and mean structural similarity index (MSSIM) on held-out billiard test simulations, as well as training time and parameter count. While deeper models with more filters offer slightly improved SSIM in some cases, we find that a relatively compact architecture with 3 × 3 kernels and two pooling stages provides a favorable balance between accuracy and computational cost. The final architecture described above is used for all results reported in this work.

## 3.5 Training and evaluation metrics

The ConvLSTM networks are trained using a mean squared error (MSE) loss between predicted and ground-truth images. Although the model predicts multiple future frames for each input sequence, the loss is evaluated only on the terminal predicted frame. For this frame, the squared difference between predicted and true RGB intensities is computed for each pixel and channel and averaged to yield a scalar loss. Optimization is performed using the Adam optimizer with minibatches of sequence pairs $(\mathbf{x}_t, \mathbf{y}_t)$ drawn from the training set.

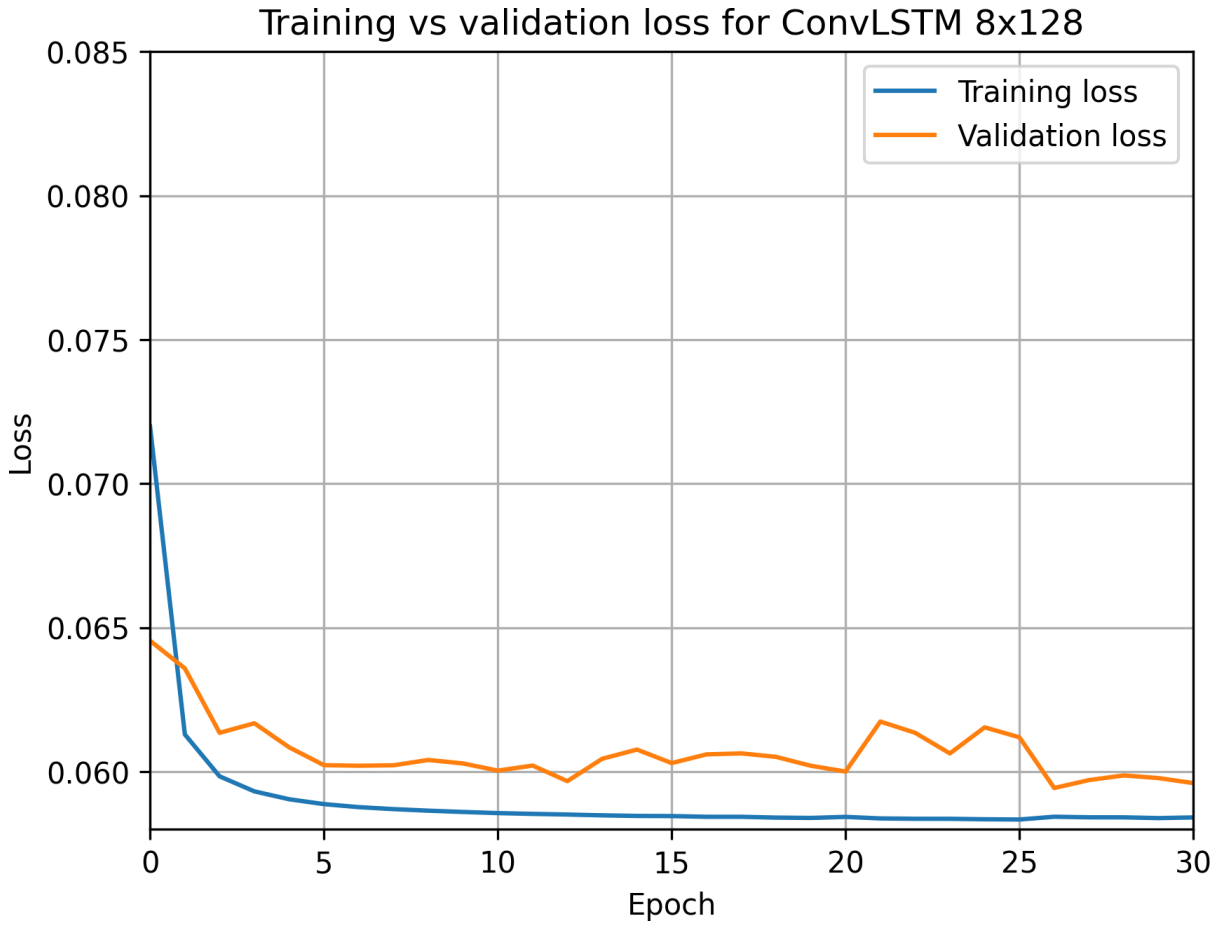

Figure 3: Training and validation loss for ConvLSTM 8x128 (3x3) model

Figure 3 shows the evolution of the training and validation loss over epochs. Both curves decrease monotonically and plateau after 30 epochs. Convergence of validation loss suggests no obvious overfitting of the ConvLSTM model.

To assess perceptual similarity between predicted and reference images, we compute the structural similarity index (SSIM) on a per-frame basis. For a given frame index $k$, we compare the predicted image $\hat{f}_k$ to the corresponding FLAG image $f_k$ and evaluate SSIM over all pixels and channels, yielding a scalar value in $[-1, 1]$ (with 1 indicating perfect agreement). In the figures, we report SSIM as a function of frame index for each test case, and in the tables we quote representative values or averages over selected time intervals. In both the billiard and granular compaction problems, SSIM provides a more informative measure of visual and structural agreement than MSE alone, as it accounts for local contrast and texture as well as overall intensity.

To identify the best ConvLSTM architecture, we conducted hyperparameter tuning by varying filter sizes, amount of filters, kernel sizes, and amount of average pooling. Six different ConvLSTM frameworks were trained on the FLAG dataset, evaluated on Mean Squared Error (MSE) and Structural Similarity Index (SSIM) [25].

SSIM is a method of comparing how similar two images are to each other by evaluating structural information losses. As opposed to error–sensitivity methods such as MSE, which purely calculate differences in pixel values between images, SSIM focuses on capturing per-ceptual quality measures of an image. In this way, SSIM evaluates differences in images with the goal of approximating how the human visual system (HVS) perceives structural differences.

SSIM is recorded on a scale from −1 to 1, where 1 indicates that the images are exactly similar, and values near zero or negative indicate large differences between the images. SSIM considers three characteristics of an image in its comparison: *luminance*, *contrast*, and *structure*.

Specifically, let $x$ and $y$ be two corresponding regions of two different images extracted from the same spatial location. Luminance $l$, contrast $c$, and structure $s$ are defined as:

$$l(x, y) = \frac{2\mu_x\mu_y}{\mu_x^2 + \mu_y^2}, \qquad c(x, y) = \frac{2\sigma_x\sigma_y}{\sigma_x^2 + \sigma_y^2}, \qquad s(x, y) = \frac{\sigma_{xy}}{\sigma_x\sigma_y},$$

where $\mu_x$, $\mu_y$, $\sigma_x^2$, $\sigma_y^2$, and $\sigma_{xy}$ denote the mean of $x$, the mean of $y$, the variance of $x$, the variance of $y$, and the covariance between $x$ and $y$, respectively.

A basic similarity index can then be defined as the product of these three components:

$$S(x, y) = l(x, y)\, c(x, y)\, s(x, y) = \frac{4\mu_x\mu_y\sigma_{xy}}{(\mu_x^2 + \mu_y^2)(\sigma_x^2 + \sigma_y^2)}.$$

However, this expression becomes unstable when the denominator approaches zero. To address this, the final SSIM index introduces stabilizing constants:

$$SSIM(x, y) = \frac{(2\mu_x\mu_y + C_1)(2\sigma_{xy} + C_2)}{(\mu_x^2 + \mu_y^2 + C_1)(\sigma_x^2 + \sigma_y^2 + C_2)}.$$

The constants $C_1$ and $C_2$ are defined as:

$$C_1 = (K_1L)^2, \qquad C_2 = (K_2L)^2,$$

where $L$ is the dynamic range of pixel values in the image ($L = 1$ for normalized RGB images, $L = 255$ for standard grayscale images), and $K_1$ and $K_2$ are small constants chosen such that $C_1$ and $C_2$ only influence the computation when the denominator is close to zero.

Table 1 below shows the MSE and MSSIM values of the six frameworks, evaluated using two FLAG test data sets that the ML framework did not see during training. The lowest MSE and highest MSSIM for each data set are highlighted in red, with the chosen framework in bold. The framework in bold was chosen as it balanced a high MSSIM value across the two data sets with training time (the framework with the highest MSSIM values had nearly double the training time despite similar MSSIM).

| **Model** | **# params** | **MSE (D1)** | **MSSIM (D1)** | **MSE (D2)** | **MSSIM (D2)** | **Train (s)** |
|---|---|---|---|---|---|---|
| ConvLSTM (3×3): 8×64, 4×AvgPool | 2,375,747 | 0.006355 | 0.958071 | 0.006226 | 0.952963 | 5367.73 |
| ConvLSTM (3×3): 4×64–4×128, 4×AvgPool | 5,536,163 | 0.005914 | 0.94914 | 0.005750 | 0.94468 | 5371.04 |
| **ConvLSTM (3×3): 8×128, 2×AvgPool** | **8,936,611** | **0.007791** | **0.98812** | **0.007676** | **0.98269** | **6456.06** |
| ConvLSTM (5×5): 4×64–4×128, 2×AvgPool | 15,371,875 | 0.0077907 | 0.988212 | 0.0076802 | 0.982773 | 12402.13 |
| ConvLSTM (5×5): 8×64, 1×AvgPool | 6,601,155 | 0.0077497 | 0.98685 | 0.007640 | 0.981050 | 5703.95 |
| ConvLSTM (5×5): 8×128, 1×AvgPool | 26,309,507 | 0.005812 | 0.91837 | 0.0058123 | 0.9097877 | 6468.81 |

Table 2: Comparison of ConvLSTM architectures on the billiard problem. For each model, we report the number of trainable parameters, mean squared error (MSE) and mean SSIM (MSSIM) on two billiard test simulations, and training time.

The layers of the encoder process the input sequence both spatially and temporally, allowing the network to extract hierarchical spatiotemporal features. Spatial resolution is progressively reduced through average pooling operations applied after the second and fourth ConvLSTM layers. These pooling steps downsample the feature maps from the initial resolution of 128×128 to 64×64, and subsequently to 32×32 pixels, thereby compressing the spatial information while preserving temporal dependencies. The output of the final encoder layer forms a latent representation of the input sequence at a spatial resolution of 32×32 and a temporal depth of 10 steps.

The decoder consists of two ConvLSTM layers, each also configured with 128 filters and ReLU activations. To reconstruct the original spatial resolution, we apply upsampling operations after each decoder layer. Specifically, the spatial dimensions are upsampled from 32×32 to 64×64 after the first decoder layer, and from 64×64 to 128×128 after the second, restoring the resolution to match the input size. The final output layer is a ConvLSTM layer with a sigmoid activation function, which produces the predicted frame sequence with pixel values normalized between 0 and 1.

This architecture is well-suited for modeling complex pressure propagation patterns captured in the training set. The ConvLSTM layers capture these dynamics by maintaining memory over time while convolving over spatial neighborhoods, allowing the model to learn how localized pressure changes propagate through space. The hierarchical structure of the encoder extracts features at multiple spatial resolutions, enabling the network to capture both fine-grained details and broader propagation patterns. The decoder then reconstructs these learned spatiotemporal representations into full-resolution predictions, effectively modeling the pressure field's evolution over time. As a result, the network can satisfactorily forecast pressure propagation in this billiards ball example.

As summarized schematically in Figure 2, the encoder–decoder ConvLSTM consists of six encoding layers and two decoding layers with intermediate pooling and upsampling.

# 4 Results

## 4.1 Billiard impact surrogate modeling

We first evaluate the ConvLSTM surrogate on the 10-ball billiard impact problem. The network is trained on ten FLAG simulations with varying cue-ball positions and velocities, as described in Section 3. Two additional simulations, corresponding to unseen combinations of position and velocity, are reserved as test cases.

Figure 4 plots SSIM as a function of frame index for several candidate architectures evaluated on the first billiard test set. SSIM is computed for each frame by comparing the ML predicted image to the corresponding FLAG image and averaging over all pixels. The curves show that the chosen architecture maintains relatively high SSIM over the prediction horizon and compares favorably to deeper and wider alternatives, while requiring fewer parameters and shorter training times.

Qualitative comparisons of predicted and reference images are shown in Figure 5. We display selected frames spanning the early, intermediate, and late stages of wave propagation. In the early frames, the ConvLSTM accurately reproduces the initial impact of the cue ball and the emergence of a localized high-pressure region at the contact. As the wave propagates through the finite-volume mesh within each ball, the surrogate captures the primary features of the pressure field, including the wavefront's shape, position, and interactions with curved boundaries. The predicted ball positions closely track the FLAG simulations, demonstrating that the ConvLSTM has learned the coupling between wave propagation and grain motion encoded in the training data.

At later times, as the pressure wave reflects from boundaries and the system approaches a more complex, multi-wave state, minor discrepancies appear in the fine details of the pressure field. However, the global structure of the wave pattern and the overall grain configuration remain well reproduced. Quantitatively, MSE values remain small and SSIM values remain high across both billiard test sets, indicating robust performance under variations of initial cue-ball position and impact velocity.

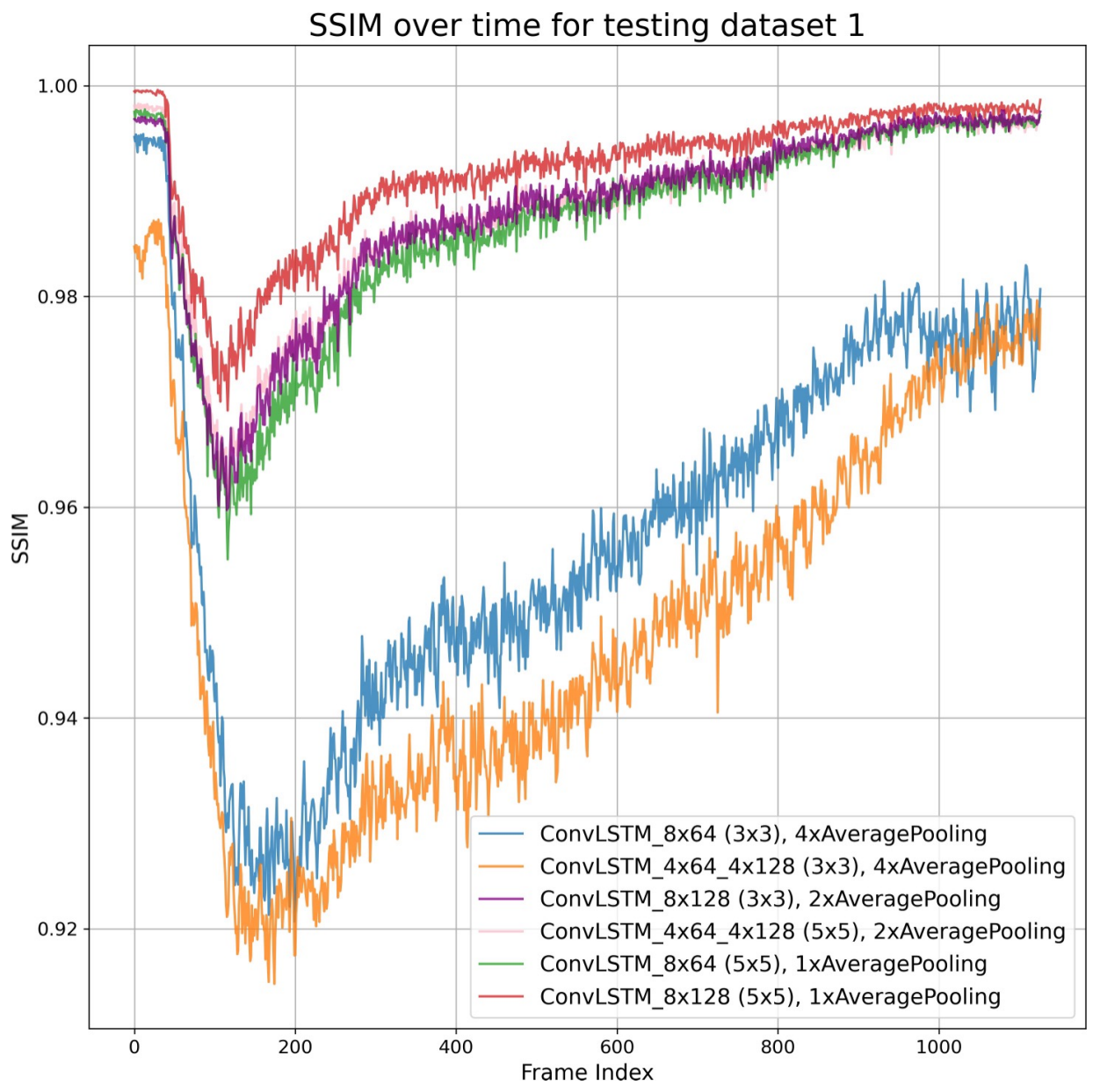

Figure 4: Structural similarity index (SSIM) versus frame index for several ConvLSTM architectures on the first billiard test set. The chosen architecture maintains high SSIM over the prediction horizon while using a moderate number of parameters.

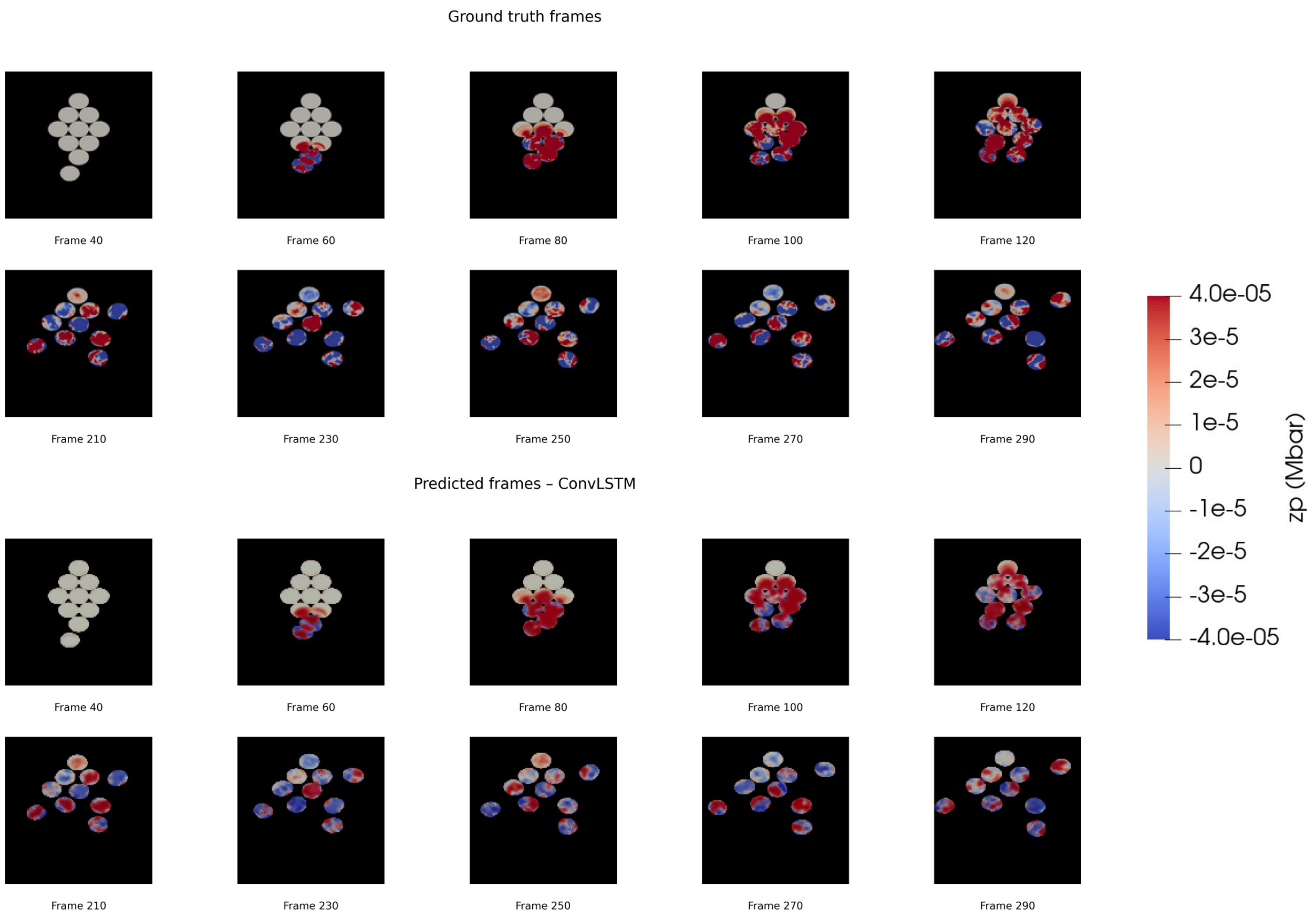

Figure 5: Comparison of ground truth (top) and ConvLSTM predictions (bottom).

## 4.2 Granular compaction: proof-of-concept extension

To assess whether the ConvLSTM framework can extend to more realistic granular configurations, we apply the same architecture to mesoscale simulations of weak shock compaction in granular salt modeled as an ensemble of circular grains. The ConvLSTM framework is trained in the same manner as for the billiard problem, using sequences of ten frames as input and predicting the subsequent ten frames. We then evaluate the trained model on the 100 m/s and 300 m/s test case simulations.

Qualitative comparisons for the 100 m/s and 300 m/s cases are shown in Figures 6 and 7, respectively. In both cases, the ConvLSTM successfully captures the *location* and *overall shape* of the compaction front and the primary direction of stress propagation. The predicted compaction front tracks the FLAG front as it advances into the granular pack, and the broad structure of the high-pressure region behind the front is reproduced.

The surrogate reproduces the front position and large-scale pressure morphology, but attenuates high-frequency pore-scale features in the compacted region—an expected behavior for convolutional sequence models trained with a pixelwise MSE objective. Once the shock has traversed a region and the grains have largely collapsed into a dense pack, the high-pressure zone appears as a more uniform band in the ConvLSTM predictions, with individual grain shapes and small pores no longer resolved as sharply as in the FLAG images. This loss of sharpness is observed for both 100 m/s and 300 m/s impacts, and is consistent with the smoothing effect expected from convolutional filters trained with an MSE loss.

The SSIM curves in Figure 8 highlight this behavior for the 100 m/s and 300 m/s test cases. For the 100 m/s case, SSIM tends to decrease over time, indicating that prediction quality gradually deteriorates as the front advances and the internal stress pattern becomes more complex. In contrast, for the 300 m/s case, SSIM initially decreases but later recovers and increases as the simulation proceeds. We interpret this as a consequence of the more uniform shock front at higher impact speeds: once the front has formed and the compaction pattern stabilizes, the pressure field becomes easier for the ConvLSTM to predict, leading to higher structural similarity. Quantitatively, the aggregated MSE and MSSIM values in Table 3 remain on the order of $10^{-3}$ and around 0.8, respectively, for both test cases.

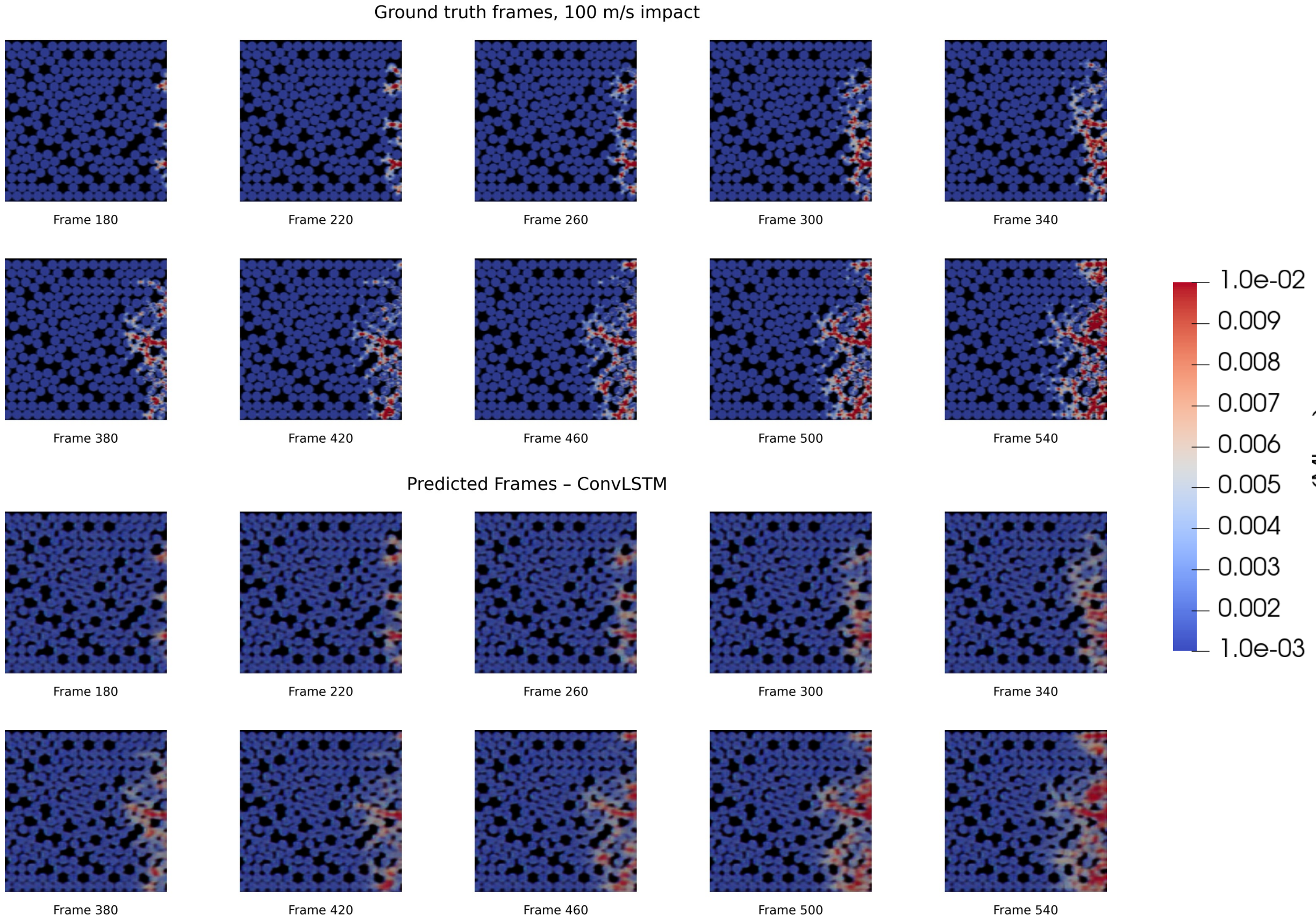

Figure 6: Comparison of FLAG (top) and ConvLSTM (bottom) pressure fields for the granular compaction test at 100 m/s, at selected frames. The ConvLSTM captures the position and overall shape of the compaction front but smooths fine pore-scale details behind the front.

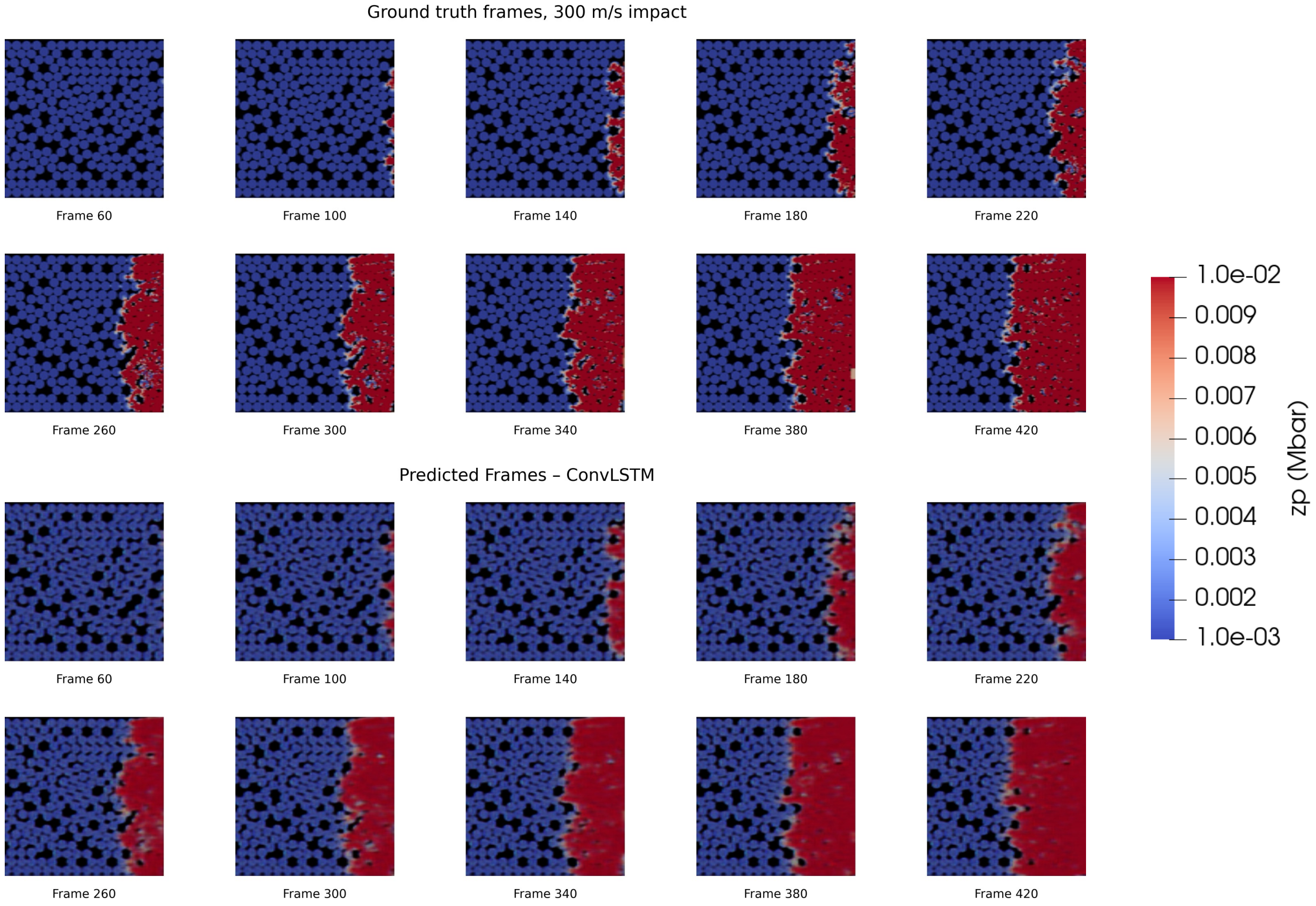

Figure 7: Comparison of FLAG (top) and ConvLSTM (bottom) pressure fields for the granular compaction test at 300 m/s. At higher impact speed the shock front is more uniform, which the ConvLSTM predicts more consistently, while still smoothing small pores in the compacted region.

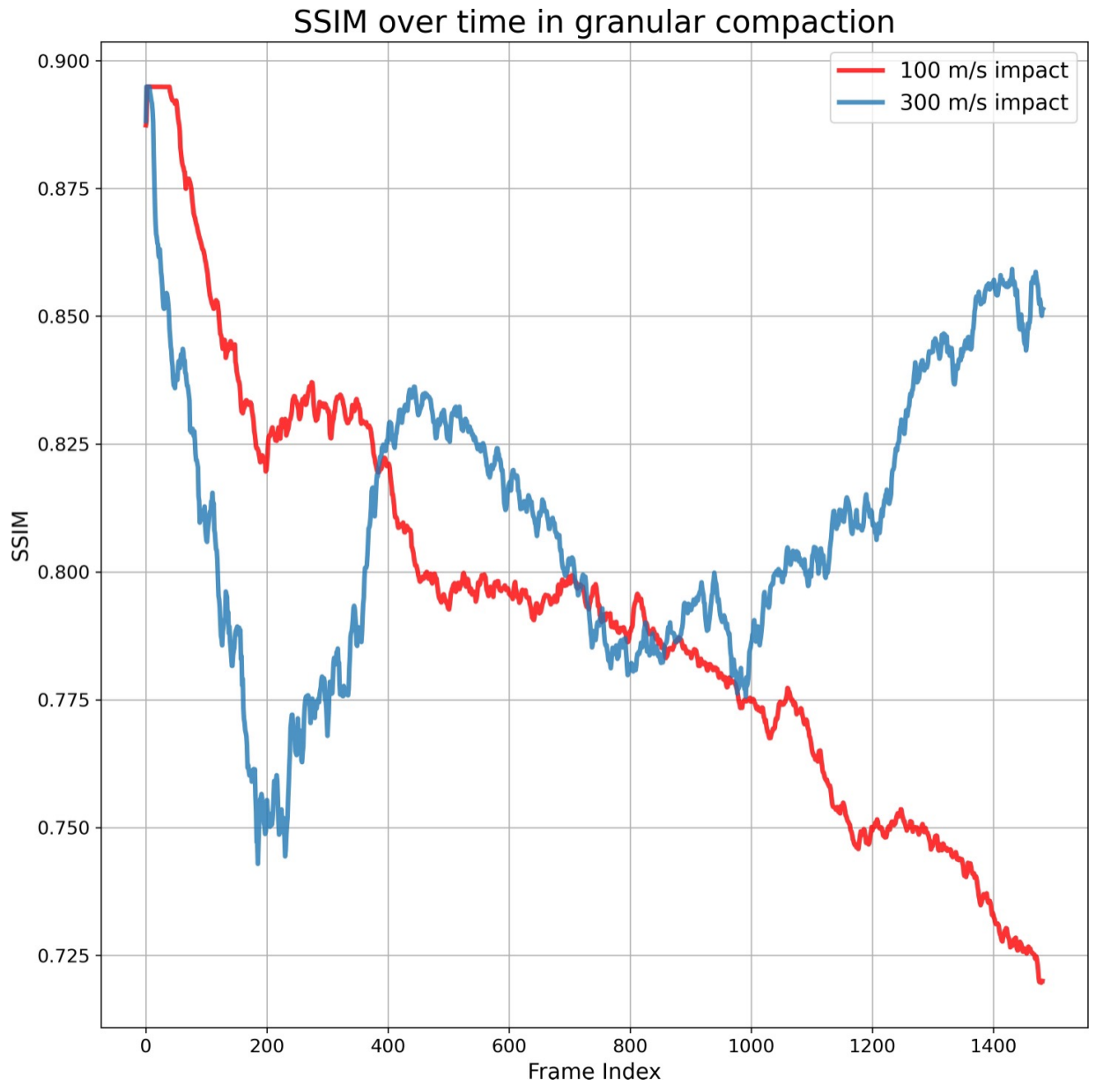

Figure 8: SSIM versus frame index for the granular compaction tests at 100 m/s and 300 m/s. SSIM decreases over time for 100 m/s, whereas for 300 m/s it initially decreases and then recovers as the compaction front becomes more uniform.

Table 3: Mean squared error (MSE) and mean SSIM (MSSIM) for the granular compaction tests at 100 m/s and 300 m/s, averaged over the prediction horizon.

| Impact speed | MSE | MSSIM |
|---|---|---|
| 100 m/s | 0.00745 | 0.793 |
| 300 m/s | 0.00467 | 0.810 |

# 5 Discussion

The results presented above demonstrate that ConvLSTM networks trained on mesoscale hydrocode simulations can serve as effective surrogates for spatiotemporal wave propagation in granular systems. In the simplified billiard problem, the surrogate accurately predicts the evolution of both the pressure field and grain positions for unseen combinations of cue-ball position and velocity. This indicates that the network has learned a representation of the underlying dynamics that generalizes across initial conditions within the training envelope.

The granular compaction problem presents a more stringent test, with thousands of interacting grains, evolving pore structures, and a propagating compaction front. Here, the ConvLSTM is able to capture the large-scale structure of the shock front and its dependence on impact speed, even when evaluated at piston speeds that were completely withheld from training. At the same time, the model struggles to reproduce fine pore-scale details behind the front, particularly in highly compacted regions where grain shapes and small pores are strongly distorted. This limitation is not surprising: the training objective penalizes pixel-wise errors uniformly, and the convolutional architecture naturally favors smooth approximations of high-frequency structures.

From a modeling perspective, these findings highlight both the promise and the current limitations of purely data-driven spatiotemporal surrogates for granular mesoscale simulations. On the positive side, ConvLSTM models can be trained to emulate complex, nonlinear dynamics directly from image sequences, without requiring explicit hand-crafted features or reduced-order representations.

On the other hand, the loss of fine-scale resolution suggests several avenues for improvement. First, the training objective could be augmented with multi-scale or perceptual losses that place greater weight on small-scale structures, or with physics-inspired regularization terms that penalize unphysical artifacts. Second, hybrid architectures that combine ConvLSTM layers with attention mechanisms or graph-based representations may offer improved capacity for capturing localized interactions between grains. Third, explicit multi-resolution training strategies could be employed, in which the network is trained to predict both coarse and fine-scale components of the pressure and configuration fields.

It is also important to emphasize that, in this work, we deliberately leverage existing, validated mesoscale simulations from Seo et al. [5] as the source of training data. This allows us to focus on the ML methodology and its ability to reproduce known mesoscale behavior, rather than re-establishing the validity of the underlying physics model. A single ConvLSTM rollout of N frames takes O(1) seconds on a GPU vs 12 hours of FLAG CPU time. Future work could combine the ConvLSTM surrogate with experimental data, for example by training on or calibrating to in situ imaging of rapid compaction [26], thereby enhancing fidelity and extending the surrogate beyond the parameter space covered by current simulations.

# 6 Conclusions

We have presented a ConvLSTM-based spatiotemporal surrogate model for mesoscale simulations of wave propagation in granular materials. Using high-fidelity FLAG hydrocode simulations as training data, we demonstrated that in a simplified 10-ball billiard impact problem, the ConvLSTM accurately predicts the evolution of pressure fields and grain positions for unseen combinations of initial cue-ball position and velocity. The same ConvLSTM architecture can be applied, without modification, to mesoscale simulations of weak shock compaction in granular salt modeled as an ensemble of circular grains, capturing the position and shape of the compaction front and its dependence on impact speed. The surrogate reproduces large-scale features of the pressure field and compaction front, but tends to smooth fine pore-scale details behind the front, especially in highly compacted regions. These results suggest that ConvLSTM networks provide a promising framework for constructing fast, image-based surrogates of mesoscale hydrocode simulations in granular mechanics. While the current implementation is purely data-driven and uses a simple MSE loss, future work may incorporate physics-informed constraints, multi-scale loss functions, or hybrid architectures to improve fidelity at small scales. Such developments would enable more accurate and efficient exploration of parameter space, uncertainty quantification, and coupling to continuum-scale models in applications involving rapid compaction and shock loading of granular materials.

# Credit authorship contribution statement

**Kathleen Martinus**: Conceptualization; Software; Methodology; Formal analysis; Validation; Visualization; Writing – original draft; Writing – reviewing & editing; Data curation. **Dawa Seo**: Software; Methodology; Formal analysis; Validation; Visualization; Writing – reviewing & editing; **Sushan Nakarmi**: Conceptualization; Supervision; Methodology; Validation; Writing – original draft; Writing – reviewing & editing. **Nitin Daphalapurkar**: Conceptualization; Supervision; Methodology; Validation; Writing – original draft; Writing – reviewing & editing; Funding acquisition.

# 7 Acknowledgments

Research on computational modeling was supported by U.S. Department of Energy's ASC/PEM/EM through LANL with Christina Scovel as the project leader and D.J. Luscher as the program manager. LANL is operated by Triad National Security, LLC, for the National Nuclear Security Administration of U.S. Department of Energy (Contract No. 89233218CNA000001).

# Data availability

The data supporting the findings of this study are available from the lead author upon reasonable request and in accordance with the policies of the Los Alamos National Laboratory.

# Declaration of generative AI and AI-assisted technologies in the manuscript preparation process

During the preparation of this work the authors used OpenAI's software package chatGPT in order to reason. After using this tool, the authors reviewed and edited the content as needed and take full responsibility for the content of the published article.